\documentclass[a4paper,12pt]{article}

\newcommand{\be}{\begin{equation}}
\newcommand{\ee}{\end{equation}}
\newcommand{\ba}{\begin{eqnarray}}
\newcommand{\ea}{\end{eqnarray}}
\usepackage[dvips]{graphicx}
\usepackage{color}

\usepackage{amssymb}

\begin{document}

{\centerline{\Large\bf Phenomenological model explaining }}
{\centerline{\Large\bf  Hubble Tension origin}}

\vspace{1cm}

{\centerline{G.S.~Bisnovatyi-Kogan$^{1,2,3}$}}

\vspace{0.5cm}

\noindent $^1$Space Research Institute, Russian Academy of Sciences, Moscow, Russia\\
$^2$National Research Nuclear University MEPhI, Moscow, Russia.

\vspace{1cm}
\noindent

\section*{Abstract}

 One of the problem revealed recently in cosmology is a so-called Hubble tension (HT), which is
 the difference between values of the present Hubble constant, measured by observation of the universe at redshift $z \lesssim 1$, and by observations of a distant universe with CMB fluctuations originated at $z \sim 1100$.
In this paper we suggest, that this discrepancy  may be explained  by deviation of the cosmological expansion from a standard Lambda-CDM
%simple Friedman
model of a flat universe, during the period after recombination at $z \lesssim 1100$, due to action of additional variable component of a dark energy of different origin..
We suppose, that a dark matter (DM) has a common origin with a variable component of a dark energy (DEV).
DE presently may have two components, one of which is the Einstein constant $\Lambda$, and another, smaller component DEV ($\Lambda_V$) comes from the remnants of a scalar fields responsible for inflation.
Due to common origin and interconnections the densities of DEV and DM are supposed to be connected, and remain almost constant during, at least, the time after recombination, when we may approximate $\rho_{DM}=\alpha \rho_{DEV}$.
This part of the dark energy in not connected with the cosmological constant $\Lambda$, but is defined by existence of scalar fields with a variable density.
Taking into account the influence of DEV on the universe expansion
we find the value of $\alpha$ which could remove the HT problem.  In order to maintain the almost constant DEV/DM energy density ratio during the time interval at $z<1100$, we suggest an existence of a wide mass DM particle distribution.

{\it Keywords} dark energy, dark matter, Hubble constant
%==================================================

\section{Introduction}

Recently a challenge in cosmology was formulated, because of different values, obtained from different experiments, of the Hubble constant at present epoch.
There is a significant discrepancy (tension) between
the {\em Planck} measurement from cosmic microwave background (CMB) anisotropy, where the best-fit model gives \cite{Ade:2015xua},\cite{planck18},

\begin{equation}\label{eq:h0_P18}
H_0^{\rm P18} = 67.36 \pm 0.54 ~~\mathrm{km \, s}^{-1} \, \mathrm{Mpc}^{-1} \,,
\end{equation}
and measurements using type Ia supernovae (SNIa)  calibrated with Cepheid
distances~\cite{1998Riess,1999Perlmutter,riess16, riess18a,riess19},

\begin{equation}\label{eq:h0_R19}
H_0^{\rm R19} = 74.03 \pm 1.42 ~~ \mathrm{km \, s}^{-1} \, \mathrm{Mpc}^{-1} \,.
\end{equation}
Measurements using time delays from lensed quasars~\cite{wong19} gave the value
$H_0 = 73.3^{+1.7}_{-1.8} \, \mathrm{km \, s}^{-1} \, \mathrm{Mpc}^{-1}$,
{while in Ref. \cite{yuan19} it was found $H_0 = 72.4 \pm1.9\, \mathrm{km \, s}^{-1} \,
\mathrm{Mpc}^{-1}$} using the tip of the red giant branch applied to SNIa, which is
independent of the Cepheid distance scale. Analysis of a compilation of these and other
recent high- and low-redshift measurements shows~\cite{Verde:2019ivm} that the discrepancy between {\it Planck} \cite{planck18}, and any three independent late-Universe measurements is between $4\sigma$ and $6\sigma$. Different sophisticated explanations for appearance of HT have been proposed  \cite{kk2016,md2018,pskk2018,yang2019,sunny2019,di2019}, see also \cite{um2015,bal2016,rossi2019,hhunter}, and
new experiments have been proposed for checking the reliability of this
tension  \cite{ben19}.

Dark matter (DM) and dark energy (DE) represent about 96\% of the universe constituents \cite{1998Riess,1999Perlmutter,2003Spergel}, but their origin is still not clear. The present value of DE density  may be represented by Einstein cosmological constant $\Lambda$ \cite{lambda}, but also may be a result of the action of the Higgs-type scalar fields, which are supposed to be the reason of the inflation in the early universe \cite{guth}, see also \cite{star,mukh,linde}. The value of  the induced $\Lambda$, suggested for the inflation,  is many orders of magnitude larger, than its present value, and  no attempts have been done, to find a connection between them. The origin of DM is even more vague.
 There are numerous suggestions for its origin \cite{dmrev}, but none of these possibilities has been experimentally or observationally confirmed, while many of them have been disproved.

 One exciting observational feature of the modern cosmology is the fact that we live
just in the period, where densities of DE and DM are comparable to each other.
The equality by the order of magnitude between DE and DM densities may take place not by chance, but
may be connected with
their common origin and evolution. If we use an analogy with electromagnetic field, we could suppose, that massive
DM particles are born by the massless
scalar field, and their energy densities remain comparable in
some periods of the universe expansion.

  In this paper, in order to explain the origin of the Hubble Tension, we introduce a %small
   variable part of the cosmological "constant" $\Lambda_V$, proportional to the matter density $\rho_{DM}=\alpha \rho_{DEV}$. This part $\Lambda_V$ influence the cosmological expansion at large redshifts, where the influence of the real Einstein constant $\Lambda$ is negligible.
  The value of $\Lambda_V$ is represented by a small component of DE, which we define as DEV.
  We suppose here, without knowledge of physical properties of DM particles, that there is a wide spectrum of DM particle, which could be produced by DEV until the present time.

  It seems necessary because at decreasing of the DEV field strength in the expanding universe it would be able to show mutual transformations only with DM particles of decreasing mass.
The existence of particles with a very low rest mass (axions \cite{axion}) is considered often as a candidate for DM.

  We consider a model of the expanding universe after recombination, at $z<\sim 1100$, with a fixed ratio $\alpha$ of energy densities between DEV, connected with a scalar field,
 and DM.  We suggest, that a birth of the ordinary matter in the
process of inflation takes place also, but  DM  is born  more effectively. If the mass spectrum of DM particles prolongs to very small masses, then we may expect almost constant DEV-DM ratio.
 In the inflation model of the universe, only a scalar field was born at the very beginning, and a matter was created in the process of expansion from the dynamic part of the scalar field density.

Here we show, that in presence of DEV the Hubble value $H$ is decreasing with time slower. This create a larger present value of $H_0$, removing the Hubble tension at $\alpha \sim 5$.

  \section{Universe with common origin of DM and DE}

%\subsection{Expanding universe with equipartition. Scalar field in the equipartition universe}

  The scalar field with the potential $V(\phi)$, $\phi$ is the intensity of the scalar field,
is considered as the main reason of the inflation \cite{guth}, but see \cite{star}. The equation for the scalar field in the expanding universe is written as \cite{peeb}

\begin{equation}
\label{eq1}
\ddot\phi+3\frac{\dot a}{a}\dot\phi=-\frac{dV}{d\phi}.
\end{equation}
Here $a$ is a scale factor in the flat expanding universe \cite{fridman}. The density $\rho_V$, and pressure $P_V$ of the scalar field {\footnote{In most equations below it is taken $c=1$.}} are defined as \cite{peeb}

\begin{equation}
\label{eq2}
\rho_V=\frac{\dot\phi^2}{2}\,+\,V, \quad    P_V= \frac{\dot\phi^2}{2}\,-\,V.
\end{equation}
Consider the universe with the initial scalar field, at initial intensity $\phi_{in}$ and initial potential $V_{in}$, and at zero derivative $\dot\phi_{in}=0$. The derivative of the scalar field intensity is growing on the initial stage of inflation.

Let us suggest, that after reaching the relation

\begin{equation}
\label{eq3}
 \dot\phi^2 = 2\alpha V,
 \end{equation}
it is preserved during farther expansion. The kinetic part of the scalar field is transforming into matter, presumably, dark matter, and the constant $\alpha$ determines the ratio of the the dark energy density, represented by $V$, to the matter density, represented by the kinetic term. As follows from observations, the main part of DE is represented presently by the Einstein constant $\Lambda$. At earlier times the input of constant $\Lambda$ is smaller, than the input of $\Lambda_V$, for wide interval of constant $\alpha$ values.

  Let us consider an expanding flat universe, described by the Friedmann equation \cite{fridman}

   \begin{equation}
\label{eq4}
\frac{\dot a^2}{a^2}=\frac{8\pi
G}{3}\rho +\frac{\Lambda}{3}.
\end{equation}
Introduce

\begin{eqnarray}
\label{eq5}
\rho_\phi=V, \quad    P_\phi= -V,
\quad \rho_m=\frac{\dot\phi^2}{2}, \quad P_m=\beta\frac{\dot\phi^2}{2},\quad {\mbox{with}} \quad \rho_m=\alpha\rho_\phi.
\end{eqnarray}
We suggest, that only part $\beta$ of the kinetic term make the input in the pressure of the matter, so it follows from (\ref{eq3}),(\ref{eq5})

\begin{equation}
\label{eq6}
\rho=\rho_\phi+\rho_m=(1+\alpha)V,\quad   P=P_\phi+P_m=-(1-\alpha\beta)V.
\end{equation}
The adiabatic condition

\begin{equation}
\label{eq7} \frac{d\rho}{\rho
+P}=-\frac{d{\cal V}}{\cal V}=-3\frac{da}{a}, \quad {\cal V}\quad {\mbox{is\,\,the \,\,volume}},
\end{equation}
may be written as

\begin{eqnarray}
\label{eq8}
\dot\rho=-3\frac{\dot a}{a}(\rho_\phi+\rho_m+ P_\phi+P_m)=-3\frac{\dot a}{a}(\rho_m+P_m)=-3\alpha\frac{1+\beta}{1+\alpha}\frac{\dot a}{a}\rho.
\end{eqnarray}

\begin{eqnarray}
\label{eq8a}
\frac{\rho}{\rho_*}=\left(\frac{a}{a_*}\right)^{\frac{3\alpha}{2(1+\alpha)}} \quad {\mbox{for}}\quad \beta=0.
\end{eqnarray}

\section{A universe without a cosmological constant $\Lambda$: a toy model}

Suggest cosmological constant $\Lambda=0$, when DE is created only by the part of the scalar field $\Lambda_V$, represented by $V$.
The expressions for the total density $\rho$, scaling factor $a$, and Hubble "constant" $H$ follow from (\ref{eq4})-(\ref{eq8}) as

\begin{eqnarray}
\label{eq9}
\frac{a}{a_*}=(6\pi G \rho_* t^2)^{\frac{1+\alpha}{3\alpha(1+\beta)}}\left(\frac{\alpha(1+\beta)}
{1+\alpha}\right)^{\frac{2(1+\alpha)}{3\alpha(1+\beta)}}
=\left(\frac{\rho_*}{\rho}\right)^{\frac{1+\alpha}{3\alpha(1+\beta)}}
=\left(\frac{t}{t_*}\right)^{\frac{2(1+\alpha)}{3\alpha(1+\beta)}}, \nonumber\\
\rho=\left(\frac{1+\alpha}{\alpha(1+\beta)}\right)^2 \frac{1}{6\pi Gt^2},\quad H=\frac{\dot a}{a}=\frac{2(1+\alpha)}{3\alpha(1+\beta)t}.\qquad\qquad
\end{eqnarray}
Here $\rho_*=\rho(t_*),\,\,\,a_*=a(t_*)$, $t_*$ is an arbitrary time moment.
Write the expressions for particular cases. For $\beta=1/3$ (radiation dominated universe) it follows from (\ref{eq9})

\begin{eqnarray}
\label{eq10}
\frac{a}{a_*}=(6\pi G \rho_* t^2)^{\frac{1+\alpha}{4\alpha}}\left[\frac{4\alpha}
{3(1+\alpha)}\right]^{\frac{1+\alpha}{2\alpha}}
=\left(\frac{\rho_*}{\rho}\right)^{\frac{1+\alpha}{4\alpha}}
=\left(\frac{t}{t_*}\right)^{\frac{1+\alpha}{2\alpha}}, \nonumber\\
\rho=\left(\frac{3(1+\alpha)}{4\alpha}\right)^2 \frac{1}{6\pi Gt^2},\quad H=\frac{\dot a}{a}=\frac{1+\alpha}{2\alpha t}.\qquad\qquad
\end{eqnarray}
For the value of $\beta=0$ (dusty universe, $z<1100$) we have

\begin{eqnarray}
\label{eq11}
\frac{a}{a_*}=(6\pi G \rho_* t^2)^{\frac{1+\alpha}{3\alpha}}\left[\frac{\alpha}
{1+\alpha}\right]^{\frac{2(1+\alpha)}{3\alpha}}
=\left(\frac{\rho_*}{\rho}\right)^{\frac{1+\alpha}{3\alpha}}
=\left(\frac{t}{t_*}\right)^{\frac{2(1+\alpha)}{3\alpha}}, \nonumber\\
\rho=\left(\frac{1+\alpha}{\alpha}\right)^2 \frac{1}{6\pi Gt^2},\quad H=\frac{\dot a}{a}=\frac{2(1+\alpha)}{3\alpha t}.\qquad\qquad
\end{eqnarray}

\subsection{Removing the Hubble Tension}

As follows from above consideration, a linear connection between densities $\rho_\phi$ and $\rho_m$ leads to a change of rate of the universe expansion. In the inverse dependence of the Hubble "constant" on time $H=\lambda/t$, the value of $\lambda$ occupies the interval $(\frac{2}{3(1+\beta)}\, < \lambda <\, \frac{8}{3(1+\beta)})$ for  $(\infty\, > \alpha >\, 1/3)$, respectively.
Decrease of the matter (DM) input into the density, in comparison with the DE, leads to increasing of the speed of the universe expansion, so that, according to (\ref{eq9}), we have formally $\lambda \rightarrow\infty$ at $\alpha\rightarrow 0$, what means that the expansion is becoming of the exponential de Sitter type at $\alpha=0$.

The main idea of removing the tension is the following.
The CMB measurements give the value of the Hubble constant $H_r$, at the redshift $z\sim 1100$, close to the moment of recombination. This value is used for calculation of the present value of $H_0$.

For analysis of the Hubble tension it is more convenient to use logarithmic variables, so that from (\ref{eq:h0_P18}),(\ref{eq:h0_R19}) (\ref{eq9}) we have
\begin{eqnarray}
\label{eq19}
\log{H_0^{P18}}=\log{67.36}=1.83;\quad \log{H_0^{R19}}=\log{74.03}=1.87; \nonumber \\ \log\frac{H_0^{R19}}{H_0^{P18}} = \Delta\log{H_0}=0.04 \qquad \qquad \qquad.
\end{eqnarray}
The Planck value $H^P_r$ was measured at the moment of recombination $z_r \approx 1100$,
and extrapolated to the present time using dusty flat Friedmann model as  \cite{fridman}

 \begin{equation}
\label{eq20}
z+1=\frac{\omega}{\omega_0}=\frac{a_0}{a}=\left(\frac{t_0}{t}\right)^{\frac{2}{3}},\quad
H=\frac{\dot a}{a}=\frac{2}{3t},\quad  \frac{H_0}{H}=\frac{t}{t_0}=(z+1)^{-\frac{3}{2}}.
\end{equation}
In the case of equipartition universe the extrapolation should be done using Eq. (\ref{eq9}).

Numerical modeling of large scale structure formation give the preference to the cold dark matter model, corresponding to $P_m\approx 0,\,\,\,\beta=0$. We suppose, that the dynamical part of scalar field give birth to dark energy matter in the form  of massive DM particles.
 In this procedure the Hubble tension is connected with incorrect extrapolation by Eq.(\ref{eq20}). From the value of Hubble tension in Eq.(\ref{eq19}) we may estimate $\alpha$ from the condition, that both measurements are correct, but the reported value ${H_0^{P18}}$ is coming from the incorrect extrapolation, and the actual present epoch value of the Hubble constant is determined by ${H_0^{R19}}$.

 \begin{eqnarray}
 \label{eq21}
\log{H^P_r}=\log{H_0^{P18}}+\frac{3}{2}\log{z_r}\approx 1.83+\frac{3}{2}\log{1101};\qquad\nonumber
\\
\log{H^P_r}=\log{H_0^{R19}}+\frac{3\alpha}{2(1+\alpha)}\log{z_r}
 \approx 1.87+\frac{3\alpha}{2(1+\alpha)}\log{1101};\\
 \frac{1}{1+\alpha}=\frac{2}{3}\frac{\log H_0^{R19}-\log H_0^{P18}}{\log 1101}
 \approx 0.0088; \quad
\alpha \approx 114. \nonumber .
\end{eqnarray}
We see here that a very small presence of a dark energy $\rho_\phi\approx 0.0088\rho_m$ could solve the HT problem in the universe without the cosmological constant $\Lambda$.

\section{A universe in presence of the cosmological constant $\Lambda$}

Equations (\ref{eq5})-(\ref{eq8}) are valid in presence of $\Lambda$. Solution of Eq.(\ref{eq4}) with nonzero $\Lambda$ is written in the form

\begin{eqnarray}
\label{eq14}
\left(\frac{a}{a_*}\right)^{\frac{3\alpha(1+\beta)}{2(1+\alpha)}}=\sqrt{\frac{8\pi G \rho_*}{\Lambda c^2}}
\sinh\left(\sqrt\frac{\Lambda}{3}\frac{3\alpha(1+\beta)}{2(1+\alpha)}ct\right)
=\sqrt{\frac{\rho_*}{\rho}};
\qquad \sqrt{\frac{\Lambda c^2}{8\pi G\rho}}\\
=\sinh\left(\sqrt\frac{\Lambda}{3}\frac{3\alpha(1+\beta)}{2(1+\alpha)}ct\right),
\quad
H=\frac{\dot a}{a}=\sqrt{\frac{\Lambda c^2}{3}}\coth\left(\sqrt\frac{\Lambda}{3}\frac{3\alpha(1+\beta)}{2(1+\alpha)}ct\right).
\end{eqnarray}
For the dusty universe $(\beta=0)$, after recombination at $z<1100$ we have

\begin{eqnarray}
\label{eq16}
\left(\frac{a}{a_*}\right)^{\frac{3\alpha}{2(1+\alpha)}}=\sqrt{\frac{8\pi G \rho_*}{\Lambda c^2}}
\sinh\left(\sqrt\frac{\Lambda}{3}\frac{3\alpha}{2(1+\alpha)}ct\right)=
\sqrt{\frac{\rho_*}{\rho}};
\qquad \sqrt{\frac{\Lambda c^2}{8\pi G\rho}}\\
=\sinh\left(\sqrt\frac{\Lambda}{3}\frac{3\alpha}{2(1+\alpha)}ct\right),
\quad
H=\frac{\dot a}{a}=\sqrt{\frac{\Lambda c^2}{3}}\coth\left(\sqrt\frac{\Lambda}{3}\frac{3\alpha}{2(1+\alpha)}ct\right).
\label{eq16b}
\end{eqnarray}
The dusty universe without DEV, is described by relations, following from (\ref{eq16}) (\ref{eq16b}) at $\alpha \rightarrow \infty$, giving

 \begin{eqnarray}
\label{eq17a}
\left(\frac{a}{a_*}\right)^{\frac{3}{2}}=\sqrt{\frac{8\pi G \rho_*}{\Lambda c^2}}
\sinh\left(\sqrt\frac{\Lambda}{3}\frac{3}{2}ct\right)=\sqrt{\frac{\rho_*}{\rho}};\qquad
\sqrt{\frac{\Lambda c^2}{8\pi G\rho}}\\
=\sinh\left(\sqrt\frac{\Lambda}{3}\frac{3}{2}ct\right),\qquad
\label{eq17b}
H=\frac{\dot a}{a}=\sqrt{\frac{\Lambda c^2}{3}}\coth\left(\sqrt\frac{\Lambda}{3}\frac{3}{2}ct\right).
\end{eqnarray}

\subsection{Removing the Hubble Tension}

From Eqs. (\ref{eq16},(\ref{eq20}) we obtain for dusty universe, at $\beta=0$, the following connection
between the recombination red shift $z_r\approx 1100$, the present age of the universe $t_0$ and the age of
 the universe $t_r$, corresponding to the recombination, in the form

 \begin{eqnarray}
\label{eq22}
z_{r\alpha}+1=
\frac{a_{*}}{a_{r\alpha}}=\left[
\sqrt{\frac{8\pi G \rho_*}{\Lambda c^2}}
\sinh\left(\sqrt\frac{\Lambda}{3}\frac{3\alpha}{2(1+\alpha)}ct_{r\alpha}\right)\right]
^{-\frac{2(1+\alpha)}{3\alpha}}\\
=\left(\frac{\rho_*}{\rho_{r\alpha}}\right)^{-\frac{(1+\alpha)}{3\alpha}}, \qquad
\label{eq16c}
H=\frac{\dot a}{a}=\sqrt{\frac{\Lambda
c^2}{3}}\coth\left(\sqrt\frac{\Lambda}{3}\frac{3\alpha}{2(1+\alpha)}ct_{r\alpha}\right).
\end{eqnarray}
 Here indices with $\alpha$ indicate the values in the universe with DEV, The indices "0" is related to the present time, at $z_0=0$. The indices "r" is related to the moment of recombination. The values in the universe
without DEV don't contain  $\alpha$ in the indices, and are written as

  \begin{eqnarray}
\label{eq23}
z_r+1=\frac{a_{*}}{a_r}=
\left[
\sqrt{\frac{8\pi G \rho_*}{\Lambda c^2}}
\sinh\left(\sqrt\frac{\Lambda}{3}\frac{3}{2}ct_r\right)\right]
^{-\frac{2}{3}}
=\left(\frac{\rho_*}{\rho_{r}}\right)^{-\frac{1}{3}},  \\
\label{eq16d}
H=\frac{\dot a}{a}=\sqrt{\frac{\Lambda
c^2}{3}}\coth\left(\sqrt\frac{\Lambda}{3}\frac{3}{2}ct_r\right).\qquad\
\end{eqnarray}
The observational data are connected with a red shift, so we find the expression for the time as a function of the red shift in the form

  \begin{eqnarray}
\label{eq23a}
ct_{r\alpha} = {\frac{2(1+\alpha)}{3\alpha}}\sqrt{\frac{3}{\Lambda}}\sinh^{-1}{\left[\sqrt{\frac{\Lambda c^2}{8\pi G \rho_*}}(z_{r\alpha}+1)^{\frac{3\alpha}{2(1+\alpha)}}\right]}\\ \approx
\frac{\alpha+1}{\alpha\sqrt{6\pi G\rho_*}}(z_{r\alpha}+1)^{\frac{3\alpha}{2(1+\alpha)}},\qquad\qquad
\nonumber.
\end{eqnarray}

  \begin{eqnarray}
\label{eq23b}
ct_{0\alpha} = {\frac{2(1+\alpha)}{3\alpha}}\sqrt{\frac{3}{\Lambda}}\sinh^{-1}{\left(\sqrt{\frac{\Lambda c^2}{8\pi G \rho_*}}\right)}.
\end{eqnarray}
At $t=t_{r\alpha}$ the argument of $"\sinh"$ is very small, so only first term in the expansion remains. Corresponding values for the universe without DEV are written as

  \begin{eqnarray}
\label{eq23c}
ct_{r} = {\frac{2}{3}}\sqrt{\frac{3}{\Lambda}}\sinh^{-1}{\left[\sqrt{\frac{\Lambda c^2}{8\pi G \rho_*}}(z_{r}+1)^{\frac{3}{2}}\right]} \approx
\frac{1}{\sqrt{6\pi G\rho_*}}(z_{r}+1)^{\frac{3}{2}},
\end{eqnarray}

  \begin{eqnarray}
\label{eq23d}
ct_{0\alpha} = {\frac{2(1+\alpha)}{3\alpha}}\sqrt{\frac{3}{\Lambda}}\sinh^{-1}{\left(\sqrt{\frac{\Lambda c^2}{8\pi G \rho_*}}\right)}.
\end{eqnarray}
For $z_{r\alpha}=z_r \equiv z_{rec}$ we obtain a connection between $t_{r\alpha}$ and $t_r$ as

  \begin{eqnarray}
\label{eq23t}
t_{r\alpha} = \frac{\alpha+1}{\alpha}(z_{rec}+1)^{\frac{3}{2(1+\alpha)}} t_r.
\end{eqnarray}

The ratio of Hubble constants at present time $t_0$ to its value at the recombination time $t_{r\alpha}$, in presence of DEV, using (\ref{eq16b}), is written as

\begin{eqnarray}
\label{eq24}
\frac{H_{0\alpha}}{H_{r\alpha}}=
 \frac{\coth\left(\sqrt\frac{\Lambda}{3}\frac{3\alpha}{2(1+\alpha)}ct_0\right)}
 {\coth\left(\sqrt\frac{\Lambda}{3}\frac{3\alpha ct_{r\alpha}}{2(1+\alpha)}\right)}=
 \frac{\tanh\left(\sqrt\frac{\Lambda}{3}\frac{3\alpha ct_{r\alpha}}{2(1+\alpha)}\right)}
 {\tanh\left(\sqrt\frac{\Lambda}{3}\frac{3\alpha}{2(1+\alpha)}ct_0\right)}=
 \frac{\sqrt\frac{\Lambda}{3}\frac{3\alpha}{2(1+\alpha)}ct_{r\alpha}}
 {\tanh\left(\sqrt\frac{\Lambda}{3}\frac{3\alpha}{2(1+\alpha)}ct_0\right)}.
\end{eqnarray}
 The corresponding values for the case without DEV, at $\alpha\rightarrow\infty$, with recombination time $t_r$ are written as

 \begin{eqnarray}
\label{eq25}
\frac{H_{0}}{H_{r}}=
 \frac{\coth\left(\sqrt\frac{\Lambda}{3}\frac{3}{2}ct_0\right)}
 {\coth\left(\sqrt\frac{\Lambda}{3}\frac{3 ct_{r}}{2}\right)}=
 \frac{\tanh\left(\sqrt\frac{\Lambda}{3}\frac{3ct_{r}}{2}\right)}
 {\tanh\left(\sqrt\frac{\Lambda}{3}\frac{3}{2}ct_0\right)}=
 \frac{\sqrt\frac{\Lambda}{3}\frac{3}{2}ct_{r}}
 {\tanh\left(\sqrt\frac{\Lambda}{3}\frac{3}{2}ct_0\right)}.
\end{eqnarray}
Using Eqs. (\ref{eq23t}),(\ref{eq24}),(\ref{eq25}), we obtain the ratio between correct value $H_{0\alpha}$, identified with the local measurement of $H$, and the value $H_{0}$, obtained by calculations without account of DEV, in the form

 \begin{eqnarray}
\label{eq27}
\frac{H_{0\alpha}}{H_{0}}=
 \frac{\tanh\left(\sqrt\frac{\Lambda}{3}\frac{3}{2}ct_0\right)}
 {\tanh\left(\sqrt\frac{\Lambda}{3}\frac{3\alpha}{2(1+\alpha)}ct_0\right)} (z_{rec}+1)^{\frac{3}{2(1+\alpha)}}.
\end{eqnarray}

\subsection{Numerical estimations}

Let us consider the following commonly accepted, averaged parameters of the universe, used in relation with HT explanation.

Local Hubble constant $H_l=73$ km/s/Mpc

Distantly measured Hubble constant $H_d=67.5$ km/s/Mpc

Total density of the flat universe $\rho_{tot}=2\cdot 10^{-29} h^2=1.066\cdot 10^{-29}$ g/cm$^3$, where

\noindent h = H/100 km/s/Mpc, with $h=0.73$

Distantly measured densities of the universe components, $\rho_{\Lambda_*}=0.7 \rho_{tot}=7.5\cdot 10^{-30}$ g/cm$^3$; $\rho_{m*}=0.3 \rho_{tot}$.

Locally measured cosmological constant density \cite{1999Perlmutter} $\rho_{\Lambda_0}=(0.44\div 0.96)\,\rho_{tot}$, (2$\sigma$ statistics)

$\Lambda=\frac{8\pi G \rho_{\Lambda_*}}{c^2}=1.40\cdot 10^{-56}$ cm$^{-2}$

Average age of the universe $t_0=4.35 \cdot 10^{17}$ seconds \cite{wik}

The ratio of two Hubble constants is the following  $HT_r=\frac{H_l}{H_d}= 1.08$

Identifying $H_{0\alpha}\equiv H_l$, $H_0\equiv H_d$, we obtain from (\ref{eq27})

\begin{eqnarray}
\label{eq28}
\frac{H_{0\alpha}}{H_{0}}=1.08=
 \frac{\tanh\left(\sqrt\frac{\Lambda}{3}\frac{3}{2}ct_0\right)}
 {\tanh\left(\sqrt\frac{\Lambda}{3}
 \frac{3\alpha}{2(1+\alpha)}ct_0\right)}(z_{rec}+1)^{\frac{3}{2(1+\alpha)}}\\
 =\frac{\tanh(1.337)}{\tanh\left(1.337\frac{\alpha}{1+\alpha}\right)}
 (1101)^{\frac{3}{2(1+\alpha)}}=
 \frac{0.87095}{\tanh\left(1.337\frac{\alpha}{1+\alpha}\right)}(1101)^{\frac{3}{2(1+\alpha)}}.
\nonumber
\end{eqnarray}
Finally we obtain the equation for $\alpha$ in the form

\begin{eqnarray}
\label{eq29}
\tanh\left(1.337\frac{\alpha}{1+\alpha}\right)=0.8064 (1101)^{\frac{3}{2(1+\alpha)}},\quad
\alpha\approx 140.
\end{eqnarray}
Taking $\rho_{mr}=0.3 \rho_{tot}$, we obtain $\rho_{DEV}=0.0022 \rho_{tot}$, and the effective dark energy density at present time should be equal to $\Omega_{0eff}=0.7022$, what is inside the limits of local measurement of $\Lambda_0$, mentioned above. The local matter density of the flat universe at present time is $\Omega_{0m}=0.2978$, leading to the dark matter density $\Omega_{0dm}=0.2578$, for the  baryon density $\Omega_{0b}=0.04$.

\bigskip

\section{Discussion}

In order to explain the origin of the Hubble Tension we have introduced a variable part of the cosmological constant $\Lambda_V$, proportional to the matter density.
We have solved the Friedmann equation in presence of the relation (\ref{eq3}), and have found the value of $\alpha$, at which HT disappeared. The present DEV density, needed for explanation of HT phenomena is very small relative to the cosmological constant $\Lambda$.
It influences the cosmological expansion at larger redshifts, where the input of the Einstein constant $\Lambda$ is small. Presently the situation is opposite,  $\Lambda\gg \Lambda_V$, because decreasing of a matter density during cosmological expansion determines the transition from the quasi-Friedmann to quasi-de Sitter stage.
The estimation of the density $\rho_{\Lambda_V}$ at present epoch corresponds to $\Omega_{\Lambda_V}\approx 0.0022$, what slightly increases the present dark energy density. In the flat universe is determines  decreasing of the dark matter density due to transfer of  its energy into DEV in the condition of "equipartition" at constant $\alpha$.

We have used for estimations a constant ratio of $\rho_{\Lambda_V}/\rho_m =1/\alpha$ for the universe expansion after recombination, at $z<1100$, but deviations from this law should not change qualitatively the conclusion, that a relatively small average contribution of the variable $\Omega_{\Lambda_V}$
may explain the difference of Hubble constant measured at local, and high $z$ distances.

 The present parameters of LCDM model, have been estimated from analysis of CMB fluctuations measurements in WMAP and PLANCK experiments, having a power-law spectrum of adiabatic scalar perturbations . The procedure is based on a search of extremes in the multidimensional parameter space. The presence of HT (if real) is adding an additional restriction in this problem. The universe parameters obtained in this process may be changed with this additional restriction. It is not possible to predict these changes without making a full set of computations within the suggested hypothesis, even for such a small DEV density. The computations could be performed in presence of a variable $\alpha$. Decreasing of dark matter leads to decreasing of the field amplitude, what may prevent the energy exchange between DM and DEV in absence of very light DM particles.
Note, that account of the cosmological constant leads to moderate increasing of the $\alpha$ value of (decreasing DEV  density), in comparison with  with the toy model without $\Lambda$.

In our model the DM should be represented by a wide mass spectrum particles, and not by a unique  mass CDM particles, what is usually considered now.  By analogy with CMB, the lowest mass of DM particles should not exceed presently the value $\sim (\Omega_{DEV}/\Omega_{CMB})^{1/4}\times kT_{CMB} \approx 7\cdot 10^{-4}$ eV, for retain a possibility of almost constant $\alpha$.

\section*{Acknowledgments}

This work was supported by the Russian Science Foundation (grant no. 18-12-00378).
The author is very grateful to O.Yu. Tsupko for valuable comments.
%==%=================%==== figs 1-1a =========%================================

\begin{figure*}
%\centering
%\includegraphics[width=7cm]{./Figures/fig1.eps}
%\includegraphics[width=7cm]{./Figures/fig1a.eps}
%\caption{iii}
\label{fig1}
\end{figure*}

%==%=================%======figs 2-2a ====%================================

\end{document}